\documentclass[amstex,showpacs,preprint]{revtex4}
\usepackage{graphicx}
\usepackage{amsmath}
\usepackage{bm}
\usepackage{amsfonts}
\usepackage{amssymb}

\begin{document}

\title{Entanglement, the quantum formalism \\ and the classical world}
\author{A.\ Matzkin}
\affiliation{Laboratoire de Physique Th\'{e}orique et Mod\'{e}lisation, CNRS
Unit\'{e} 8089, Universit\'{e} de Cergy-Pontoise,\\ 95302 Cergy-Pontoise cedex, France}

\begin{abstract}
75 years after the term "entanglement" was coined to a peculiar feature
inherent to quantum systems, the connection between quantum and classical
mechanics remains an open problem. Drawing on recent results obtained in
semiclassical systems, we discuss here the fate of entanglement in a closed
system as Planck's constant becomes vanishingly small. In that case the
generation of entanglement in a quantum system is perfectly reproduced by
properly defined correlations of the corresponding classical system. We
speculate on what these results could imply regarding the status of
entanglement and of the ensuing quantum correlations.
\end{abstract}

\maketitle

\section{Introduction}

Schr\"{o}dinger coined the term "\emph{entanglement}" in 1935 \cite{s1,s2} to
describe the fact that when two systems interacted, the resulting state could
"\emph{no longer be described in the same way as before, viz. by endowing each
of them with a representative} [ie vector state in Hilbert space] \emph{of its
own}. [...] \emph{By the interaction the two representatives have become
entangled}" \cite{s2}. And he added in the same first paragraph of
Ref.\ \cite{s2} the now celebrated phrase: "\emph{I would not call that one
but rather the characteristic trait of quantum mechanics, the one that
enforces its entire departure from classical lines of thought}". And indeed,
there is a-priori nothing like entanglement in classical mechanics. Our aim in
the present note is to give some results on the properties of entanglement as
the classical limit is approached and make a few remarks on the fate of
entanglement in the classical world.

The understanding of entanglement has made some progress since 1935, in
particular these last 20 years with the advent of quantum information related
works. However these works deal essentially with qubits (a two state system
for which the action -- the spin -- is of the order of $\hslash$) relevant to
investigate only some of the conceptual aspects (quantum logic, separability,
communication constraints) that lie at the root of entanglement. Other
aspects, dealing with the quantum-classical correspondence and the emergence
of classical mechanics, call for additional tools. These are to be found in
semiclassical physics.

In short semiclassical physics \cite{brack,gutz} investigates the properties
of quantum systems by making an explicit link with the properties of the
corresponding classical system. In general, the existence of a corresponding
classical system is guaranteed by canonical quantization (the classical and
quantum Hamiltonians have the same functional dependence on phase-space
variables and operators respectively), and the quantum system behaves
semiclassically provided the actions $S_{i}$ of the system are huge relative
to Planck's constant, ie $S_{i}/\hslash\rightarrow\infty$, or in short
$\hslash\rightarrow0$. In other cases the corresponding classical system is
obtained from the first order expansion of the path integral form of the
evolution operator. In both situations the validity of the quantum-classical
correspondence hinges on the fact that as $\hslash\rightarrow0$ the
wavefunction propagates (simultaneously) along all the available trajectories
of the corresponding classical system.

Given that entanglement is a characteristic quantum property, it could appear
at first sight far-fetched to look for any quantum-classical correspondence in
entangled semiclassical systems. This is why we will first give a brief
overview of the vast amount of works that have studied the effect of the
classical underlying dynamics on the entanglement evolution. Then we will
summarize some results concerning entanglement generation that are obtained
for a scaling system whose dynamics is invariant as $\hslash\rightarrow0$,
giving rise to an apparent paradox: entanglement increases as the classical
limit is approached, but the amount of entanglement is captured with
increasing accuracy by probabilities obtained from the corresponding classical
system. We will close with some remarks regarding entanglement in the
classical world, the role of decoherence, and the status of the present
quantum mechanical formalism.

\section{Entanglement in semiclassical systems}

Assume two particles, each endowed with its own dynamics obtained from a
single particle Hamiltonian $H_{i}$ with $i=1,2$ become coupled at some time
$t=0$ via an inter-particle potential term $V_{12}.$ Classically, the
equations of motion are obtained from the total Hamiltonian
\begin{equation}
H=H_{1}+H_{2}+V_{12}. \label{5}%
\end{equation}
The individual particle trajectories are coupled by the $V_{12}$ term, and if
one resorts to a statistical description, the typical distributions that can
be defined for the entire system are given by summing correlated single
particle distributions.

Moving to the quantum case, assume each single particle system described by
(the now quantized) $H_{i}$ is in the semiclassical regime, ie the
wavefunction propagates along the classical trajectories of the classical
single particle Hamiltonian; as a result the dynamical and statistical
properties of the quantum system can be obtained from the properties of the
classical trajectories, in particular from the classical periodic orbits (see
Refs. \cite{gutz,brack} and Sec.\ 3 of \cite{sphp} for a short exposition).
When the systems are coupled, the total system wavefunction $\left\vert
\psi(t)\right\rangle $ is built from the eigenstates of the Hamiltonian
(\ref{5}); it is expressed over the product Hilbert space basis $\mathcal{H}%
=\mathcal{H}_{1}\otimes\mathcal{H}_{2}$. The change brought by the
inter-particle interaction in the dynamics of the individual particles can be
followed by computing the reduced density matrices. Hence for example from the
total density matrix%
\begin{equation}
\rho(t)=\left\vert \psi(t)\right\rangle \left\langle \psi(t)\right\vert
\end{equation}
the reduced density matrix $\rho_{1}(t)$ giving particle 1's properties is
obtained by averaging over particle 2's possible outcomes
\begin{equation}
\rho_{1}(t)=\text{Tr}_{2}\rho(t). \label{8}%
\end{equation}

Generically $\left\vert \psi(t)\right\rangle $ will be entangled.\ Even if
initially $\left\vert \psi(t=0)\right\rangle $ is chosen as a product state,
entanglement will build up during the ensuing unitary evolution. To quantify
entanglement it is customary to employ the linear entropy
\begin{equation}
\Omega(t)=1-\text{Tr}_{1}\rho_{1}^{2}(t) \label{10}%
\end{equation}
which is easier to compute than the Von Neumann entropy. $\Omega(t)$ vanishes
for product states and takes its highest value for maximally entangled states.
Note that $\Omega$ is symmetric, ie Tr$_{1}\rho_{1}^{2}=$Tr$_{2}\rho_{2}^{2}$.

Given our original assumption regarding the semiclassical regime, in the
uncoupled case $\rho_{i}(t)$ is given by a sum of classical
amplitudes.\ It makes sense to expect that the $\rho_{i}(t)$ still follow to
some extent the quantum-classical correspondence even in the presence of
$V_{12}$ (and especially so if the coupling is weak). Then we see from Eq.
(\ref{10}) that the quantum-classical correspondence will transpire in the
generation of entanglement. This observation prompted several works involving
mostly numerical and sometimes analytical approaches (it is impossible to cite
all these works here; see eg \cite{angelo99,laksha
01,znidaric03,gong03,fujisaki04,epl06,chung 09,kicked11} and Refs. therein).
One of the main issues concerns the relationship between classical chaos and
entanglement: grounded on the general idea that classically chaos enhances the
diffusion in phase space, it seems natural to expect that quantum systems with
a classically chaotic counterpart will entangle more efficiently than those
having a classical counterpart displaying regular dynamics.

Things were not so simple however (counterexamples were readily
obtained).\ Briefly stated, what matters is that the classical distributions
corresponding to the initial uncoupled quantum states dynamically evolve so as
to mix significantly.\ Then in the quantum system this will correspond to
mixtures of the probability amplitudes, yielding a non-diagonal density
matrix. The type of classical flow leading to such a situation does not need
to depend on the dynamical regimes of the uncoupled systems, but rather on the
coupling parameters and on the choice of initial distributions. This is why it
was suggested \cite{pra06} that a relevant comparison of entanglement
generation with the underlying classical dynamics should be carried out in
systems in which the coupling interaction $V_{12}$ that generates the
entanglement is the one that drives the classical dynamical regime (ie that
creates chaos). Particularly interesting systems are those in which $V_{12}$
depends on a parameter $k$ that can be varied so that for certain values of
$k$ the classical dynamics is regular, but as $k$ is varied the dynamics
becomes of the mixed phase-space type, or chaotic. An illustration concerning
such a system is given in Fig.\ 1.

\begin{figure}[tb]
\includegraphics[height=5cm]{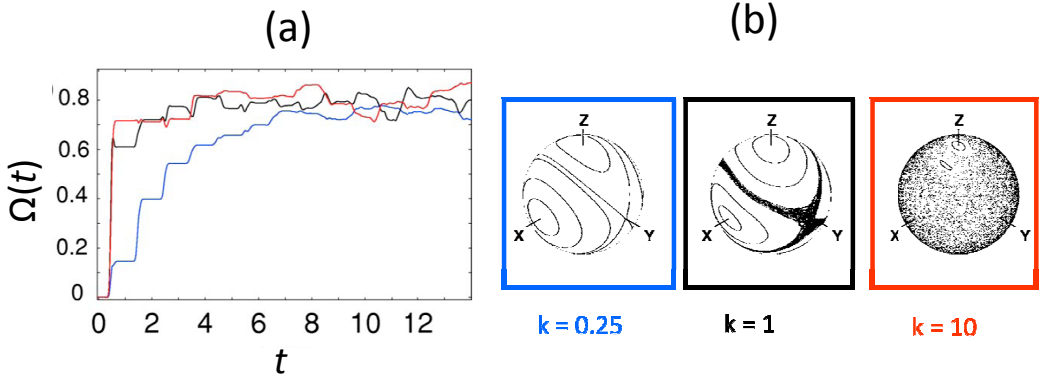}\caption{(a) The linear entropy $\Omega(t)$
gives the entanglement generation rate for a bipartite system uncoupled at $t=0$ ($t$ is
given in terms of the number of times the particles interact through $V_{12}$).
Each curve corresponds to a different value of $k$, characterizing the strength
of the contact interaction $V_{12}$ . (b) The surface of section for the corresponding classical system
is shown for different values of $k$. For $k=0.25$ the classical system displays regular dynamics (the
entanglement rate in the corresponding quantum system is shown in blue in (a)). For $k=1$ ($k=10$)
the system has mixed phase-space (chaotic) dynamics and the corresponding curves are shown in black (red) in
(a).}%
\label{figT4}%
\end{figure}

\section{Entanglement as $\hslash\rightarrow0$}

\subsection{Entanglement and classical distributions}

From the example shown in Fig.\ 1, it can be remarked that regular or chaotic
dynamics in the classical system lead to a comparable entanglement rate.
Generically however, it is true that chaos tends to translate into more
entanglement in arbitrary situations.\ It is nevertheless difficult to make
universal statements: since there is no classical quantity corresponding to
entanglement, it is not possible to compute a well-defined classical version
of Eq.\ (\ref{10}).\ Moreover for typical systems that have been investigated
numerically, $\hslash$ is still far from being negligible (relative to the
actions of the systems), since one must keep in mind that the size of the
Hilbert space (and hence of the quantum computations) increase with the
actions and the computations become therefore untractable.

Except maybe in a weak coupling regime, where exact semiclassical expansions
can be obtained analytically it is hardly possible in definite systems to
explain the behaviour of entanglement in terms of the dynamics of classical
distributions. Only very general arguments can be given, for example when the
corresponding classical dynamics is regular the entanglement behaviour is seen
to display regular oscillations (due to coherent revivals of the quantum
states constrained to remain in regular structures -- the torii), whereas this
is not the case in the presence of classically chaotic dynamics. In principle
a purely formal equivalent of Eq. (\ref{10}) can be defined \cite{gong03} by
replacing $\rho_{1}$ or $\rho_{2}$ by classical distributions and the trace by
phase-space integrals. But the linear entropy thus defined is not symmetric
(ie, the quantum relation Tr$_{1}\rho_{1}^{2}=$Tr$_{2}\rho_{2}^{2}$ is not
necessarily verified with phase-space integrals) and hardly has a physical
meaning within classical mechanics. Our strategy that we pursue below is to employ a quantum system in
which the generation of entanglement is due to a contact interaction
producing in the classical counterpart identified changes in the motions in
each of the particles. A nice property of the system is that it scales with $\hslash$,
meaning that the dynamics stays \emph{constant} as $\hslash$ is
decreased.\ This allows to effectively investigate entanglement for a given
dynamics as $\hslash\rightarrow0$.

\subsection{Scaling system}

In a classical system the action $S$ is the quantity having the dimension of
$\hslash$. In the semiclassical approximation the quantum wavefunctions take
the generic form
\begin{equation}
\psi\thicksim\sum_{t}\left\vert \det\frac{-\partial^{2}S(\mathbf{q}_{t}%
)}{\partial q^{i}\partial q_{0}^{j}}\right\vert ^{1/2}\exp iS(\mathbf{q}%
_{t})/\hslash\label{20}%
\end{equation}
for a single particle, where $t$ runs over the classical trajectories
reaching $\mathbf{q}$ from the initial point $\mathbf{q}_{0}$.\ In general, if
 one increases $S$ so that $\hslash/S\rightarrow0$,
the dynamics (both classical and quantum) is modified. For a two-particle
system the action is separable only at $t=0$ but can often be written as
$S=\sum_{i}S_{i}+\sum_{ij}S_{ij}$ (where the indices run on the particles
and/or on the degrees of freedom). Now as $\hslash/S\rightarrow0$ the dynamics
as well as the entanglement properties in the quantum system will be modified.
If the system scales however, we can modifiy the parameters and dynamical
variables of the system so that $S_{i}\rightarrow\tilde{S}_{i}/\kappa$,
$S_{ij}\rightarrow\tilde{S}_{ij}/\kappa$ (and thus $S\rightarrow\tilde
{S}/\kappa)$ and $q^{i}\rightarrow q^{i}\kappa^{\gamma}$ where $\kappa$ is a
constant. As can be seen from Eq. (\ref{20}) this scaling is tantamount to
keeping the action and the dynamics constant but rescaling an effective
Planck's constant defined by $\hslash_{\text{eff}}=\kappa\hslash$, so that by
choosing smaller values of $\kappa$ one can effectively investigate
entanglement as $\hslash\rightarrow0$.

We have recently investigated entanglement as $\hslash\rightarrow0$ in a
two-particle scaling system \cite{entang11}.\ In short, bipartite entanglement
is generated by repeated inelastic scattering of two particles -- a light
structureless particle and a heavy rotating particle, modeled by a symmetric
top with angular momentum $N$ and energy $E_{N}\propto N(N+1)$. The scattering
potential $V_{12}$ is taken to be a contact interaction so that the light
incoming particle receives a kick when it hits the rotating top. To account
for repeated scattering we add an attractive field between both particles.
Labelling $\left\vert F_{N}\right\rangle $ and $\left\vert N\right\rangle $
the quantum states of the light and heavy particles respectively ($\left\vert
F\right\rangle $ depends implicitly on $N$ because the total energy and the total angular
momentum of the entire system are conserved) a typical quantum state takes the
form%
\begin{equation}
\left\vert \psi\right\rangle =\sum_{N}B_{N}\left\vert F_{N}\right\rangle
\left\vert N\right\rangle , \label{25}%
\end{equation}
showing entanglement between the rotational state of the symmetric top and and
the energy and angular momentum of the light particle (the $B_{N}$ are just
coefficients depending on the scattering matrix elements). Initially both
particles are uncoupled, the state being $\left\vert F_{0}\right\rangle
\left\vert N_{0}\right\rangle $.

The classical version of the system hinges on employing the semiclassical link
between the deflection angle $\phi$ produced on the motion of the light
particle by the kick and the quantum scattering matrix.\ According to this
link, the kick strength can be parameterized by a coupling strength $k$, each
value of $k$ corresponding to a different quantum scattering matrix. The
scaling involves the angular momenta (which are actions) of the light and
heavy particles and the radial action of the light particle. The dynamics
remains invariant provided the orbital period of the light particle and the
rotational period of the heavy particle are adjusted accordingly.

\subsection{Entanglement and classical probabilities}

A first result \cite{entang11} concerns a simple scaling formula for the
linear entropy (\ref{10}) quantifying entanglement. Assume two values of the
effective Planck constant with $\tilde{\hslash}_{\mathrm{eff}}<\hslash
_{\mathrm{eff}}.\ $Then%
\begin{equation}
\tilde{\Omega}(t)=1-\frac{\tilde{\hslash}_{\mathrm{eff}}}{\hslash
_{\mathrm{eff}}}\left(  1-\Omega(t)\right)  , \label{simplef}%
\end{equation}
so that entanglement increases as $\hslash_{\text{eff}}\rightarrow0.$ This is
to be expected since the size of the Hilbert space increases with decreasing
$\hslash_{\mathrm{eff}}$.

The second more surprising result was to show that the linear entropy can be
given to a good approximation by computing the weights of the evolving
classical distributions. Indeed, the initial quantum density matrix
$\left\vert F_{0}\right\rangle \left\vert N_{0}\right\rangle \left\langle
F_{0}\right\vert \left\langle N_{0}\right\vert $ has a straightforward
classical counterpart (a classical distribution). This distribution evolves
encompassing several values of the classical angular momentum $N$. If one
divides classical phase-space into $q$ cells, each cell corresponding to the
volume occupied by a quantum state $\left\vert N\right\rangle \left\langle
N\right\vert $ ($q$ being the total number of quantum states), then the
classical distribution spreads across a certain number of such cells.\ By
simply counting the relative fraction of the classical distribution in each
cell, we define the probabilities $p_{N}^{cl}(t),$ corresponding to the
probability of having the classical system in which the top has an angular
momentum $N\pm\Delta_{N}/2$ (and the light particle the relevant angular
momentum and energy as imposed by the conservation laws).\ $\Delta_{N}$ is a
measure of the width of the classical cell.

The probabilities $p_{N}^{cl}(t)$ are asymptotically close to the weights of
the reduced density matrices, ie letting $\rho_{1}$ of Eq. (\ref{8}) refer to
the reduced density matrix of the rotating top, then%
\begin{equation}
\rho_{1}(t)=\sum_{N}p_{N}(t)\left\vert N\right\rangle \left\langle
N\right\vert
\end{equation}
with $p_{N}(t)\approx p_{N}^{cl}(t)$. This is due to the fact that the
off-diagonal elements of the quantum scattering matrix oscillate wildly as
$\hslash\rightarrow0,$ so that the interference terms resulting from the
incoherent sum of thousands of terms tends to vanish. Classically the quantity%
\begin{equation}
M(t)=1-\sum_{N}\left[  p_{N}^{cl}(t)\right]  ^{2}%
\end{equation}
is the mutual information quantifying the amount of mixing among the different
cells of the classical system, each cell being defined by the symmetric top
having a mean rotation number $N$ and the light particle having the
corresponding mean energy. An illustration of this behaviour is given in Fig.\ 2.

\begin{figure}[tb]
\includegraphics[height=3cm]{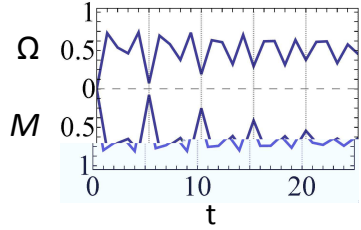}\caption{
The top panel shows the entanglement rate for the bipartite quantum system (the time is given by the number
of contact interactions between the two particles). The bottom panel shows the evolution of the mutual information
(regarding the value of $N$) of the corresponding classical system. $M$ is obtained
from the evolution of classical statistical distributions.}%
\label{ff}%
\end{figure}

\section{Entanglement, the quantum formalism and the classical world}

\subsection{Entanglement \emph{effectively} disappears in the classical limit}

The apparent paradox is that as the classical limit is approached,
entanglement increases but can be obtained from classical quantities. In other
words, the quantum system has a total non-diagonal density matrix readily
obtained from Eq. (\ref{25}), but looking at a single subsystem (whose
properties are coined in its reduced density matrix) the situation is
identical to the one that would follow if the total system was given by a
diagonal density matrix%
\begin{equation}
\rho^{cl}(t)=\sum_{N}p_{N}^{cl}(t)\left\vert F_{N}\right\rangle \left\langle
F_{N}\right\vert \otimes\left\vert N\right\rangle \left\langle N\right\vert .
\label{30}%
\end{equation}
Here the $cl$ superscript refers both to classical \emph{correlations }%
($\rho^{cl}$ is a convex combination of orthogonal projectors) and to
classical \emph{dynamics} (the $p_{N}^{cl}$ are probabilities obtained from
the classical system). In principle, the only way to distinguish $\rho$ from
$\rho^{cl}$ would involve measuring two-particle observables. But in practice,
as $\hslash\rightarrow0$ the coherences (in the \textquotedblleft pointer
basis\textquotedblright) of typical two-particle observables would yield
interference patterns with vanishing (and therefore undetectable) wavelengths,
at any rate smaller than the size of an elementary particle \cite{ballentine}.

The upshot is that at the end of the day, entanglement persists in the
classical world at a formal level but is devoid of any physical meaning: the
quantum system \emph{effectively} evolves according to Eq. (\ref{30}), that is
as a classical system. The situation is similar to the one encountered in
environmental decoherence, where it can be seen that by coupling an entangled
system to an environment, the reduced density matrix of the system $\rho_{S}$
obtained by averaging over the environment states behaves classically (in the
sense of correlations) for all practical purposes.\ This is so because the
non-diagonal terms of $\rho_{S}$ are strongly suppressed while the total
density matrix accounting for the coupled system and environment remains
entangled. What we have shown here is that for semiclassical systems there is
no need to introduce an environment: the individual components of the
entangled system behave effectively according to the laws of classical mechanics.

The point that remains to be discussed is to what extent an \emph{effective}
solution ('for all practical purposes') can be claimed to be a satisfactory
solution. Indeed, the system appears to behave classically, though formally it
remains entangled (and all the more so as the classical limit is approached).
It is well-known \cite{leggett,adler} that environmental decoherence is not a
real solution that accounts for the appearance of the classical world, unless
one assumes there was no problem to begin with (and unless one discards
multiple universe interpretations, which has other problems on its own). The
situation is similar here: there is no problem in accounting for the
appearance of a classical behaviour for entangled semiclassical systems in the
classical limit provided one assumes there was no problem to begin with. The
phrase 'no problem to begin with' is connected to the status given to the
quantum formalism with regard to reality.

\subsection{Reality and the quantum formalism}

Crudely speaking, there are two kinds of terms in our physical theories: some
theoretical terms refer to something ``out there'' in the Universe (they are
ontological, or referring terms) while other terms are purely knowledge
related (epistemic terms) \cite{matz realism}. If we take classical mechanics
as a paradigm, we encounter referring terms: the velocity, the position, the
applied forces refer to particles and fields that, paraphrasing Popper, we can
kick and that can kick us back \cite{popper}. On the other hand classical
mechanics also contains epistemic, non-referring terms: the Lagrangian or the
action encode the entire dynamical information of a given system. They live in
an abstract multi-dimensional configuration space and we can certainly not
kick them. The determination of which (if any) theoretical terms ascribe
reference is not an arbitrary choice that we can freely make (this is what
does not allow the Bohmian model to be considered as a realist account of quantum phenomena \cite{sphp}.\ We need
observational warrants in order to capture the ontological features of a
theory, and these only emerge by combining different types of experiments, observations and
logical inference.

The standard formalism of quantum mechanics does not allow to ascribe
reference in an undisputed and unambiguous manner. It is impossible to propose
a testable ontology out of the formalism as it stands today, and only weak
statements about what the theoretical terms could refer to can be made (like
the eigenvalue-eigenstate link, or the existence of invariant quantities like
the mass, the charge...). The only consistent interpretation taking the
theoretical terms of the standard formalism (like the wavefunction) at face
value as exisiting in reality would need to rely on some form of the many
worlds interpretation.

In this context, a prudent attitude would consist in endorsing the epistemic
view. If we assume the wavefunction is an epistemic term encoding information
about the system then we do not need to commit ourselves to solutions
involving objective processes that would distinguish two density matrices that
are different in nature ($\rho$ is entangled while $\rho^{cl}$ is not diagonal
in the "pointer" basis) but nevertheless give exactly the same predictions
(this, for all practical purposes, impossible by definition!). In short if
$\rho$ appears as formally entangled but in the classical limit the
entanglement cannot be detected, then from an epistemic point of view it is
correct to claim that entanglement has vanished. This viewpoint is
particularly consistent from a semiclassical perspective \cite{sphp} because
the basic quantum theoretical entity (the wavefunction) appears as being built
from classical quantities having a non-referring, \emph{epistemic} status (as
is obvious from Eq.\ (\ref{20})).

There is no need to say that this situation is hardly satisfactory.\ But it
leads us to speculate on whether the problems regarding the meaning and the
interpretation of the quantum mechanical features, such as entanglement, and
their behaviour when studying the quantum-classical transition, is not to be
found in trying to ascribe reference to the theoretical terms of the standard
formalism. At the classical level it would not make sense to build the
ontology for classical mechanics out of the Hamilton-Jacobi formalism. And the
semiclassical approach shows that by construction the quantum mechanical
quantities tend when $\hslash\rightarrow0$ to represent classical statistical
distributions expressed in the Hamilton-Jacobi formalism. While it is true
that logically nothing impedes that an apparently non-referring formalism in
one theory ends up referring to something real (and there are historical
examples that could support this assertion), it is noteworthy that 75 years
after the advent of entanglement, its nature still remains elusive.

\end{document}